 %
 %
\input harvmac.tex  
 %
\catcode`@=11
\def\rlx{\relax\leavevmode}                  
 %
 %
 %
\font\tenmib=cmmib10
\font\sevenmib=cmmib10 at 7pt 
\font\fivemib=cmmib10 at 5pt  
\font\tenbsy=cmbsy10
\font\sevenbsy=cmbsy10 at 7pt 
\font\fivebsy=cmbsy10 at 5pt  
\def\BMfont{\textfont0\tenbf \scriptfont0\sevenbf
                              \scriptscriptfont0\fivebf
            \textfont1\tenmib \scriptfont1\sevenmib
                               \scriptscriptfont1\fivemib
            \textfont2\tenbsy \scriptfont2\sevenbsy
                               \scriptscriptfont2\fivebsy}
\def\BM#1{\rlx\ifmmode\mathchoice
                      {\hbox{$\BMfont#1$}}
                      {\hbox{$\BMfont#1$}}
                      {\hbox{$\scriptstyle\BMfont#1$}}
                      {\hbox{$\scriptscriptstyle\BMfont#1$}}
                 \else{$\BMfont#1$}\fi}
 %
 %
 %
 %
\def\inbar{\vrule height1.5ex width.4pt depth0pt}
\def\sinbar{\vrule height1ex width.35pt depth0pt}
\def\ssinbar{\vrule height.7ex width.3pt depth0pt}
\font\cmss=cmss10
\font\cmsss=cmss10 at 7pt
\def\ZZ{\rlx\leavevmode
             \ifmmode\mathchoice
                    {\hbox{\cmss Z\kern-.4em Z}}
                    {\hbox{\cmss Z\kern-.4em Z}}
                    {\lower.9pt\hbox{\cmsss Z\kern-.36em Z}}
                    {\lower1.2pt\hbox{\cmsss Z\kern-.36em Z}}
               \else{\cmss Z\kern-.4em Z}\fi}
\def\Ik{\rlx{\rm I\kern-.18em k}}  
\def\IC{\rlx\leavevmode
             \ifmmode\mathchoice
                    {\hbox{\kern.33em\inbar\kern-.3em{\rm C}}}
                    {\hbox{\kern.33em\inbar\kern-.3em{\rm C}}}
                    {\hbox{\kern.28em\sinbar\kern-.25em{\sevenrm C}}}
                    {\hbox{\kern.25em\ssinbar\kern-.22em{\fiverm C}}}
             \else{\hbox{\kern.3em\inbar\kern-.3em{\rm C}}}\fi}
\def\IP{\rlx{\rm I\kern-.18em P}}
\def\IR{\rlx{\rm I\kern-.18em R}}
\def\Ione{\rlx{\rm 1\kern-2.7pt l}}
 %
 %

 %

\def\intem#1{\par\leavevmode%
              \llap{\hbox to\parindent{\hss{#1}\hfill~}}\ignorespaces}
 %


 %
\newskip\humongous \humongous=0pt plus 1000pt minus 1000pt   
\def\caja{\mathsurround=0pt}
\newif\ifdtup
 %
\def\eqalign#1{\,\vcenter{\openup2\jot \caja
     \ialign{\strut \hfil$\displaystyle{##}$&$
      \displaystyle{{}##}$\hfil\crcr#1\crcr}}\,}
 %

 %
\def\panorama{\global\dtuptrue \openup2\jot \caja
     \everycr{\noalign{\ifdtup \global\dtupfalse
      \vskip-\lineskiplimit \vskip\normallineskiplimit
      \else \penalty\interdisplaylinepenalty \fi}}}
 %
\def\eqalignno#1{\panorama \tabskip=\humongous
     \halign to\displaywidth{\hfil$\displaystyle{##}$
      \tabskip=0pt&$\displaystyle{{}##}$\hfil
       \tabskip=\humongous&\llap{$##$}\tabskip=0pt\crcr#1\crcr}}
 %

 %

 %

 %
 %
 %
 %
   \let\SS=\S       
\def\,{\hskip1.5pt}           
 %
\let\a=\alpha
\let\b=\beta
\let\c=\chi
\let\d=\delta                    
\let\e=\epsilon     
\let\f=\phi                       
                                    
\let\h=\eta

\let\j=\psi                                      
\let\k=\kappa
\let\l=\lambda                                   
\let\m=\mu
\let\n=\nu
\let\p=\pi                         
\let\q=\theta                   
         
\let\s=\sigma                   \let\S=\Sigma

\let\w=\omega

 %
 %
\def\Box{\sqcap\llap{$\sqcup$}}
\def\lapp{\lower.4ex\hbox{\rlap{$\sim$}} \raise.4ex\hbox{$<$}}
\def\gapp{\lower.4ex\hbox{\rlap{$\sim$}} \raise.4ex\hbox{$>$}}
\def\con{\ifmmode\raise.1ex\hbox{\bf*}
          \else\raise.1ex\hbox{\bf*}\fi}
\def\bo{{\raise.15ex\hbox{\large$\Box\kern-.39em$}}}
\let\iff=\leftrightarrow

\def\dual{\relax\leavevmode\lower.9ex\hbox{\titlerms*}}
\def\define{\buildrel\rm def\over =}
\let\id=\equiv
\let\8=\otimes
 %
 %
 %
 %

\let\2=\underline

 %
\def\dt#1{{\buildrel{\smash{\lower1pt\hbox{.}}}\over{#1}}}
\def\pd#1#2{{\partial#1\over\partial#2}}

\font\eightrm=cmr8
\def\6(#1){\relax\leavevmode\hbox{\eightrm(}#1\hbox{\eightrm)}}
\def\0#1{\relax\ifmmode\mathaccent"7017{#1}     
                \else\accent23#1\relax\fi}      
\def\7#1#2{{\mathop{\null#2}\limits^{#1}}}      
\def\5#1#2{{\mathop{\null#2}\limits_{#1}}}      
 %
\def\bra#1{\left\langle #1\right|}
\def\ket#1{\left| #1\right\rangle}
\def\V#1{\langle#1\rangle}

 %

 %

 %

 %
\newbox\t@b@x
\def\rightarrowfill{$\m@th \mathord- \mkern-6mu
     \cleaders\hbox{$\mkern-2mu \mathord- \mkern-2mu$}\hfill
      \mkern-6mu \mathord\rightarrow$}
\def\tooo#1{\setbox\t@b@x=\hbox{$\scriptstyle#1$}%
             \mathrel{\mathop{\hbox to\wd\t@b@x{\rightarrowfill}}%
              \limits^{#1}}\,}
\def\leftarrowfill{$\m@th \mathord\leftarrow \mkern-6mu
     \cleaders\hbox{$\mkern-2mu \mathord- \mkern-2mu$}\hfill
      \mkern-6mu \mathord-$}
\def\froo#1{\setbox\t@b@x=\hbox{$\scriptstyle#1$}%
             \mathrel{\mathop{\hbox to\wd\t@b@x{\leftarrowfill}}%
              \limits^{#1}}\,}
 %
\def\frac#1#2{{#1\over#2}}
\def\frc#1#2{\relax\ifmmode{\textstyle{#1\over#2}} 
                    \else$#1\over#2$\fi}           
\def\inv#1{\frc{1}{#1}}                            
 %
\def\Claim#1#2#3{\bigskip\begingroup%
                  \xdef #1{\secsym\the\meqno}%
                   \writedef{#1\leftbracket#1}%
                    \global\advance\meqno by1\wrlabeL#1%
                     \noindent{\bf#2}\,#1{}\,:~\sl#3\vskip1mm\endgroup}

\def\QED{\rlx\hfill$\Box$\kern-7pt\raise3pt\hbox{$\surd$}\bigskip}
 %
 %

 %
\def\muthstrut{\vphantom1}
\def\mutrix#1{\null\,\vcenter{\normalbaselines\m@th
        \ialign{\hfil$##$\hfil&&~\hfil$##$\hfill\crcr
            \muthstrut\crcr\noalign{\kern-\baselineskip}
            #1\crcr\muthstrut\crcr\noalign{\kern-\baselineskip}}}\,}

 %
\def\YT#1#2{\vcenter{\hbox{\vbox{\baselineskip0pt\parskip=\medskipamount%
             \def\Box{$\sqcap\llap{$\sqcup$}$\kern-1.2pt}%
              \def\Z{\hfil\vskip-5.8pt}\lineskiplimit0pt\lineskip0pt%
               \setbox0=\hbox{#1}\hsize\wd0\parindent=0pt#2}\,}}}
\def\EU{\rlx\ifmmode \c_{{}_E} \else$\c_{{}_E}$\fi}
\def\TM{\rlx\ifmmode {\cal T_M} \else$\cal T_M$\fi}
\def\TW{\rlx\ifmmode {\cal T_W} \else$\cal T_W$\fi}
\def\CM{\rlx\ifmmode {\cal T\rlap{\bf*}\!\!_M}
             \else$\cal T\rlap{\bf*}\!\!_M$\fi}
\def\hm#1#2{\rlx\ifmmode H^{#1}({\cal M},{#2})
                 \else$H^{#1}({\cal M},{#2})$\fi}
\def\CP#1{\rlx\ifmmode\IP^{#1}\else\IP$^{#1}$\fi}
\def\cP#1{\rlx\ifmmode\IC{\rm P}^{#1}\else$\IC{\rm P}^{#1}$\fi}

\def\sll#1{\rlx\rlap{\,\raise1pt\hbox{/}}{#1}}
\def\Sll#1{\rlx\rlap{\,\kern.6pt\raise1pt\hbox{/}}{#1}\kern-.6pt}

\let\ttt=\textstyle

 %
 %
\def\ie{\hbox{\it i.e.}}        

\def\CY{Calabi-\kern-.2em Yau}

\def\3{\ifmmode\ldots\else$\ldots$\fi}
\def\Z{\hfil\break\rlx\hbox{}\quad}
\def\3{\ifmmode\ldots\else$\ldots$\fi}
\def\?{d\kern-.3em\raise.64ex\hbox{-}}           
\def\9{\raise.43ex\hbox{-}\kern-.37em D}         
\def\ping{\nobreak\par\centerline{---$\circ$---}\goodbreak\bigskip}
 %
 %

 %

 %

\def\NP#1{{\it Nucl.\,Phys.\,}{\bf#1\,}}

\def\CMP#1{{\it Commun.\,Math.\,Phys.\,}{\bf#1\,}}

\def\IJMP#1{{\it Int.\,J.\,Mod.\,Phys.\,}{\bf#1\,}}
 %
 %
 %
\baselineskip=13.0861pt plus2pt minus1pt
\parskip=\medskipamount
\let\ft=\foot
\noblackbox
\def\SaveTimber{\abovedisplayskip=1.5ex plus.3ex minus.5ex
                \belowdisplayskip=1.5ex plus.3ex minus.5ex
                \abovedisplayshortskip=.2ex plus.2ex minus.4ex
                \belowdisplayshortskip=1.5ex plus.2ex minus.4ex
                \baselineskip=12pt plus1pt minus.5pt
 \parskip=\smallskipamount
 \def\ft##1{\unskip\,\begingroup\footskip9pt plus1pt minus1pt\setbox%
             \strutbox=\hbox{\vrule height6pt depth4.5pt width0pt}%
              \global\advance\ftno by1\footnote{$^{\the\ftno)}$}{##1}%
               \endgroup}}
 %
\def\Afour{\ifx\answ\bigans
            \hsize=16.5truecm\vsize=24.7truecm
             \else
              \hsize=24.7truecm\vsize=16.5truecm
               \fi}
\catcode`@=12
 %
 %
\SaveTimber
\def\rd{{\rm d}}
\def\jb{{\bar\j}}
\def\mb{{\bar\mu}}
\def\nb{{\bar\nu}}
\def\\{\hfil\break}
\def\Str{{\hbox{$\star\mkern-9mu\circ$}}}
 %
 %
\footline{\ifnum\pageno=1{}\else{\hss\tenrm--\,\folio\,--\hss}\fi}
\noblackbox
\rightline{hep-th/yymmddd}
\vglue+45mm
 \centerline{\titlerm   A Fermionic Hodge Star Operator}      \vskip20pt
 \centerline{\titlerms  Alfred Davis \ and \
                        Tristan H\"ubsch\footnote{$^{\spadesuit}$}
       {On leave from the Institut Rudjer Bo\v{s}kovi\'c, Zagreb,
        Croatia.\\ Supported by the US Department of Energy grant
        DE-FG02-94ER-40854.}}                                 \vskip+1mm
 \centerline{Department of Physics and Astronomy}             \vskip-1pt
 \centerline{Howard University, Washington, DC 20059}         \vskip-1pt
 \centerline{\tt thubsch\,@\,howard.edu}
\vfill

\centerline{ABSTRACT}\vskip2mm
\vbox{\rightskip=4.9em\leftskip=\rightskip\baselineskip=12pt\noindent
A fermionic analogue of the Hodge star operation is shown to have an
explicit operator representation in models with fermions, in spacetimes of
any dimension. This operator realizes a conjugation (pairing) not used
explicitly in field-theory, and induces a metric in the space of
wave-function(al)s just as in exterior calculus. If made real (Hermitian),
this induced metric turns out to be identical to the standard one
constructed using Hermitian conjugation; the utility of the induced {\it
complex\/} bilinear form remains unclear.}

\Date{September 1998\hfill}
\vfill\eject\pageno=1
\footline{\hss\tenrm--\,\folio\,--\hss}
 %
 %
\lref\rCec{S.~Cecotti: N=2 Landau-Ginzburg vs. \CY\ $\s$-models:
       Non-Perturbative Aspects. \IJMP{A6}(1991)1749--1813, Geometry
       of N=2 Landau-Ginzburg Families. \NP{B355}(1991)755--775.}

\lref\rSIF{I.B.~Frenkel, H.~Garland and G.J.~Zuckerman:
       Semi-Infinite Cohomology and String Theory. {\it
       Proc.\,Natl.\,Acad.\,Sci.\,USA\,\bf83}(1986)8442.}

\lref\rCGP{S.~Cecotti, L.~Girardello and A.~Pasquinucci:
       Non-Perturbative Aspects and Exact Results for the N=2
       Landau-Ginzburg Models. \NP{B328}(1989)701--722, Singularity
       Theory and N=2 Supersymmetry. \IJMP{A6}(1991)2427.}

\lref\rGrHa{P.~Griffiths and J.~Harris: {\it Principles of Algebraic
       Geometry}\Z (John Wiley, New York, 1978).}

\lref\rBeast{T.~H\"ubsch: {\it \CY\ Manifolds---A Bestiary for
       Physicists}\Z (World Scientific, Singapore, 2nd ed., 1994).}

\lref\rSpetz{A.~Strominger: Special Geometry.
       \CMP{133}(1990)163--180.}

\lref\rSuSyM{E.~Witten: Supersymmetry and Morse Theory.
      {\it J.\,Diff.\,Geom.\,\bf17}(1982)661--692.}

\lref\rPhases{E.~Witten, Nucl. Phys. B403 (1993) 159.}

 %
 %
\newsec{Introduction, Results and Synopsis}\noindent
Among models which contain fermionic degrees of freedom, we consider here
the simplest general type with $n$ spin-$\inv2$ variables: $\j_+^i$. The
canonically conjugate momentum of $\j_+^i$ is again a spinor which we here
denote by $\j_-^i$, and the pair satisfies the (equal time) canonical
anticommutation relations\ft{Extensions to field theory (where the
canonical variables depend on spatial coordinates in addition to time) may
be implemented by inserting appropriate  integrations over the space-like
coordinates with every summation over the variables' indices.
Alternatively, one may extend the formulae here to span over the
infinitely many fermionic creation and annihilation (\ie, Fourier
expansion mode) operators, relating then to the formalism of semi-infinite
forms~\rSIF.}:
\eqn\eCAR{ \big\{\, \j_-^i \,,\, \j_+^j \,\big\} ~=~ \d^{ij}~,\qquad
 \big\{\,\j_-^i\,,\,\j_-^j\,\big\}~=~0~
 =~\big\{\,\j_+^i\,,\,\j_+^j\,\big\}~.}
The bosonic degrees of freedom, $\f^a$, may be regarded as maps from
the (space)time (=domain) manifold into the field space (=target) manifold.
In particle physics 4-dimensional models, for example, the domain manifold
is the 3+1-dimensional Minkowski spacetime and the target manifold simply
$\IC^n$; in superstring models, the domain manifold is a Riemann surface
and the target manifold is the (10-dimensional) spacetime. Supersymmetry
exchanges bosonic and fermionic degrees of freedom, pairing
$\f^i\iff\j_+^i$, and has a number of very important consequences. Our
main result herein however will not depend on such symmetry or even
weather the number of $\f$'s equals the number of $\j_+$'s.

In coordinate representation, the wave-function(al)s of course must depend
on the canonical coordinates, \ie, on the $\f$'s and the $\j_+$'s. The
latter being nilpotent, $(\j_+^i)^2=0$, any wave-function(al) can be
expanded into a finite multinomial series in the $\j_+$'s, where the
coefficients are function(al)s of the $\f$'s. Without loss of generality
then, we may consider wave-function(al)s of a fixed fermion number, \ie,
fixed degree multinomials in the $\j_+$'s.

Geometrically, the bosonic variables provide local coordinates of
(maps into) a general manifold $X$. Owing to~\eCAR, however, the fermionic
variables are nilpotent\ft{...except in characteristic-2; in any case,
however, we assume characteristic-0 throughout.}, $(\j_\pm)^2{\id}0$, and
so span a vector space $V$. The wave-function(al)s are then simply
functions over $(X,V)$, and owing to the nilpotency of the $\j_+$'s,
elements of the sheaf ${\cal O}_X(\wedge^*V)$. Supersymmetry then enforces
$V=T_\f(X)$; again, this is unimportant for our main result.

Explicitly, we will be concerned with super wave-function(al)s of the form
\eqn\eWFn{ \ket{\w;m} ~=~
 \sum_{i_1{\cdots}i_m} \inv{m!}\,\w_{i_1{\cdots}i_m}(\f)\>
  \j_+^{i_1}{\cdots}\j_+^{i_m}\ket{0}~, }
where the vacuum $\ket{0}$ has been chosen so that
\eqn\eVac{ \j_-^i\ket{0} ~\id~0~. }
This explains the $\pm$ subscript: the $\j_+$'s act as raising operators,
while the $\j_-$'s are the annihilation operators\ft{We caution the Reader
that our present use of subscripts $\pm$ are unrelated, in principle, to
helicity or spacetime motion---as in, e.g., right- or left-mover in
2-dimensional theories.}.

Owing to the formal similarity
\eqn\eXXX{ \j_+^i \sim \rd z^i~, \qquad
           \j_-^i \sim \rd|_i \define \rd z^i\pd{}{z^i}
            \mkern20mu\hbox{(no sum on $i$)}~, }
the super wave-function(al) $\ket{\w;m}$ is analogous to an $m$-form:
\eqn\eFrm{ \w_{(m)} \define
 \w_{i_1{\cdots}i_m}\rd z^{i_1}{\wedge}{\cdots}{\wedge}\rd z^{i_m}~. }
This formal analogy has been successfully used in the by now standard use
of exterior forms and cohomology in field
theory~\refs{\rSuSyM,\rCGP,\rCec}. This makes it possible to reinterpret
and apply field theory results in differential (and algebraic) geometry,
and also the other way around.

Somewhat surprisingly, not all fermionic analogues of the general results
from exterior calculus seem to have been identified so far. In particular,
on any $n$-dimensional Riemannian manifold $X$, there exists a Hodge star
operation, $*$, which satisfies the following axioms:
\eqna\eAxi
 $$\eqalignno{
 * A^m &\to A^{n-m}~,                      &\eAxi{a}\cr
 **\w_{(m)} &= (-1)^{m(n-m)}\w_{(m)}~,     &\eAxi{b}\cr
 *(c_1\a+c_2\b) &= c_1(*\a) + c_2(*\b)~,   &\eAxi{c}\cr
 \a\wedge*\b &= \b\wedge*\a~,              &\eAxi{d}\cr
 \a\wedge*\a = 0~~~&\Rightarrow~~~\a\id0~. &\eAxi{e}\cr
 }$$
Here, $A^m$ is the space of $m$-forms, $\w_{(m)}$ as defined in
Eq.~\eFrm; $c_1,c_2$ are arbitrary real constants and $\a,\b$ are any two
exterior forms of the same degree. The linearity axiom~\eAxi{c} holds
for linear combinations of forms of different degrees. The symmetry
axiom~\eAxi{d}, however, cannot possibly hold for forms of different
degrees since the degree of the r.h.s.\ is $\deg(\a){+}n{-}\deg(\b)$,
whereas the degree of the l.h.s.\ is $\deg(\b){+}n{-}\deg(\a)$.

Finally, note that the symmetry and the non-degeneracy axioms ensure the
existence of an induced metric over the space of forms:
\eqn\eXXX{ g(\a,\b) ~\define~ \int_X \a\wedge*\b~. }

The purpose of this note is to analyze the fermionic equivalent of the
$*$-operation on $\wedge^*V$ (on $\wedge^*T_X$ when supersymmetry is
present) in physics models with fermions. This turns out to be related
but not equal to some known operations, such as Hermitian conjugation. In
addition and {\it unlike in exterior calculus}, this fermionic $*$-like
operation turns out to have an {\it explicit operator representation}.

Without much ado then, we proceed with a definition of this explicit
$*$-like operator and prove that it complies with the axioms~\eAxi{}. Once
this is done for the most rudimentary case of a model with a finite number
of (real) fermionic degrees of freedom, we consider generalizations as
appropriate for field theories and/or complex fermions.

\newsec{The Super-Hodge Star and its Action}\noindent
Begin with real theories, \ie, restrict to the case where all the variables
and all considered functions thereof are real.
 We then define
\eqn\eStar{ \star \define \sum_{p=0}^n f_p \e_{i_1{\cdots}i_n}\,
 \j_+^{i_1}{\cdots}\j_+^{i_p}\,\j_-^{i_{p+1}}{\cdots}\j_-^{i_n}~, }
where the coefficient $f_p$ will be determined by requiring $\star$ to
satisfy the axioms~\eAxi{}. No such explicit expression is known in
exterior calculus.

We will need the following iterative consequences of Eqs.~\eCAR, \eVac\ and
the antisymmetry of $\w_{i_1{\cdots}i_m}$:
\eqn\eAnn{{\eqalign{
 \j_-^i\,\w_{j_1{\cdots}j_m}\j_+^{j_1}{\cdots}\j_+^{j_m}\ket{0}
 &=\w_{j_1{\cdots}j_m}\big(\d^{ij_1}-\j_+^{j_1}\j_-^i\big)\,
                          \j_+^{j_2}{\cdots}\j_+^{j_m}\ket{0}~,\cr
 &=m\d^{ij_1}\w_{j_1{\cdots}j_m}\j_+^{j_2}{\cdots}\j_+^{j_m}\ket{0}~.\cr
   }}}
The result follows on permuting $\j_-^i$ all the way to the right, using
the antisymmetry of $\w_{j_1{\cdots}j_m}$ to combine the $m$ terms with
$m{-}1$ $\j_+$'s; the final term, with $\j_-^i$ all the way to the
right, vanishes on account of Eq.~\eVac. Repeating this $(n{-}p)$ times and
labeling the $\j_-$'s with some forethought, we obtain
\eqn\eRec{ \j_-^{i_{p+1}}{\cdots}\j_-^{i_n}\,\ket{\w;m} ~=~
     {1\over[m{-}(n{-}p)]!}\, \d^{i_nj_1}{\cdots}\d^{i_{p+1}j_{n-p}}\,
        \w_{j_1{\cdots}j_m}\j_+^{j_{n-p+1}}{\cdots}\j_+^{j_m}\ket{0}~, }
where the right-hand side vanishes for $m{<}(n{-}p)$ since then
${1\over[m{-}(n{-}p)]!}\id0$.

\subsec{The degree axiom, Eq.~\eAxi{a}}\noindent
We then calculate:
\eqna\eStr
 $$\eqalignno{ \star\ket{\w;m}
 &= \sum_{p=0}^nf_p\e_{i_1{\cdots}i_n}\j_+^{i_1}{\cdots}\j_+^{i_p}
    \j_-^{i_{p+1}}{\cdots}\j_-^{i_n}\ket{\w;m}~,            &\eStr{a}\cr
 &\buildrel{\eRec}\over{=}
 \sum_{p=0}^n{f_p\over[m{-}(n{-}p)]!}
    \, \d^{i_nj_1}{\cdots}\d^{i_{p+1}j_{n-p}}\w_{j_1{\cdots}j_m}
     \e_{i_1{\cdots}i_n}\cr
 &\mkern100mu{\times}\underbrace{\j_+^{i_1}{\cdots}\j_+^{i_p}
        \j_+^{j_{n-p+1}}{\cdots}\j_+^{j_m}}_{2p+m-n\define q}\ket{0}~.
 &\eStr{b}\cr
 }$$
Next, we use the identities
\eqna\eIds
 $$\eqalignno{ \j_+^{i_1}{\cdots}\j_+^{i_q}
 &\id {1\over q!}\d^{i_1{\cdots}i_q}_{j_1{\cdots}j_q}
                               \j_+^{j_1}{\cdots}\j_+^{j_q}~, &\eIds{a} \cr
 \d^{i_1{\cdots}i_q}_{j_1{\cdots}j_q}
 &\id {1\over (n{-}q)!}
   \e^{i_1{\cdots}i_qj_{q+1}{\cdots}j_n}\e_{j_1{\cdots}j_n}~, &\eIds{b} \cr
 }$$
to obtain (with $q=2p{+}m{-}n$)
 $$\eqalignno{ \star\ket{\w;m}
 &=\sum_{p=0}^n{f_p\over[m{-}(n{-}p)]!}
    \, \d^{i_nj_1}{\cdots}\d^{i_{p+1}j_{n-p}}\w_{j_1{\cdots}j_m}
     \e_{i_1{\cdots}i_n} \cr
 &\mkern40mu\times {1\over q!(n{-}q)!}
  \e^{i_1{\cdots}i_pj_{n-p+1}{\cdots}j_mk_{q+1}{\cdots}k_n}
   \e_{\ell_1{\cdots}\ell_qk_{q+1}{\cdots}k_n}
    \j_+^{\ell_1}{\cdots}\j_+^{\ell_q}\ket{0}~,
 &\eStr{c}\cr
 &=\sum_{p=0}^n{f_p\>p!\over[m{-}(n{-}p)]!q!(n{-}q)!}
    \, \d^{i_nj_1}{\cdots}\d^{i_{p+1}j_{n-p}}\w_{j_1{\cdots}j_m}\cr
 &\mkern60mu\times
  \d^{j_{n-p+1}{\cdots}j_mk_{q+1}{\cdots}k_n}
    _{i_{p+1}~{\cdots}~i_q~i_{q+1}{\cdots}i_n}
      \e_{k_1{\cdots}k_n}\,\j_+^{k_1}{\cdots}\j_+^{k_q}\ket{0}~.
 &\eStr{d}\cr
 }$$

As ${1\over r!}\id0$ for $r<0$, the summands above are nonzero only for
$(m{+}p{-}n),q,(n{-}q)\geq0$. These conditions, respectively, imply
\eqn\eXXX{ p\geq n{-}m~,\qquad
           p\geq \inv2(n{-}m)~,\qquad
           p\leq \inv2(2n{-}m)~.\qquad}
Since $n\geq m$, the second inequality may be omitted, being implied by
the first. The marginal case $p=n{-}m$ being special, we isolate
this term and find:
\eqn\eStActO{ {\eqalign{\star\ket{\w;m}
 &= f_{n{-}m}\e_{k_1{\cdots}k_n}\d^{k_nj_1}{\cdots}\d^{k_{n-m+1}j_m}
\inv{m!}\,\w_{j_1{\cdots}j_m}\,\j_+^{k_1}{\cdots}\j_+^{k_{n-m}}\ket{0}\cr
 &\mkern20mu+\sum_{p=n-m+1}^{[n-{1\over2}m]}
  {f_p\>p!\over[m{-}(n{-}p)]!q!(n{-}q)!}\cr
 &\mkern40mu\times
      \underbrace{\d^{j_{n-p+1}{\cdots}j_mk_{q+1}{\cdots}k_n}
                _{i_{p+1}~{\cdots}~i_q~i_{q+1}{\cdots}i_n}
              \d^{i_nj_1}{\cdots}\d^{i_{p+1}j_{n-p}}
               \w_{j_1{\cdots}j_m}}_{\id0}\,
           \e_{k_1{\cdots}k_n}\,\j_+^{k_1}{\cdots}\j_+^{k_q}\ket{0}~,
 }}}
where $[x]$ indicates the integer part of $x$. The indicated contractions
vanish whenever $p{>}n{-}m$, since then at least one $\d^{ij}$ is
completely contracted with $\w_{k_1{\cdots}\k_m}$. That is, the indicated
term is proportional to $\d^{j_1j_2}\w_{j_1j_2{\cdots}j_m}$, and this
vanishes identically since $\d^{ij}$ is symmetric while
$\w_{j_1{\cdots}j_m}$ is totally antisymmetric.

Thus, we have obtained that
\eqn\eStAct{{\eqalign{
 \star\ket{\w;m}&=\inv{(n{-}m)!}\,\w^\star_{i_1{\cdots}i_{n-m}}
             \j_+^{i_1}{\cdots}\j_+^{i_{n-m}}\ket{0}~,\cr
 \w^\star_{i_1{\cdots}i_{n-m}}&={\ttt{(n{-}m)!\over m!}}
             f_{n{-}m}\,\e_{i_1{\cdots}i_n}
              \d^{i_nj_1}{\cdots}\d^{i_{n-m+1}j_m}\w_{j_1{\cdots}j_m}~,\cr 
 }}}
proving that the action of the operator $\star$ on the wave-function(al)s
$\ket{\w;m}$, defined as in Eq.~\eWFn, satisfies axiom~\eAxi{a}.

\subsec{The involution axiom, Eq.~\eAxi{b}}\noindent
To fix the value of the coefficients $f_m$ in Eq.~\eStar, we calculate
\eqna\eInv
 $$\eqalignno{ \star\star\ket{\w;m}
 &=\star\Big(\inv{(n{-}m)!}\,\w^\star_{i_1{\cdots}i_{n-m}}\,
           \j_+^{i_1}{\cdots}\j_+^{i_{n-m}}\ket{0}\Big)~,     &\eInv{a}\cr
 &= f_m\e_{i_1{\cdots}i_n}\d^{i_nj_1}{\cdots}\d^{i_{m+1}j_{n-m}}
      \big[\inv{(n{-}m)!}\,\w^\star_{j_1{\cdots}j_{n-m}}\big]
      \j_+^{i_1}{\cdots}\j_+^{i_m}\ket{0}~,
                                                              &\eInv{b}\cr
 &= \frc{f_m\,f_{n-m}}{m!}
 \e_{i_1{\cdots}i_n}\d^{i_nj_1}{\cdots}\d^{i_{m+1}j_{n-m}}
  \e_{j_1{\cdots}j_n}\d^{j_nk_1}{\cdots}\d^{j_{n-m+1}k_m}\cr
 &\mkern60mu\times
    \w_{k_1{\cdots}k_m}\,\j_+^{i_1}{\cdots}\j_+^{i_m}\ket{0}~,&\eInv{c}\cr
 &= \frc{f_m\,f_{n-m}}{m!}
    \e_{i_1{\cdots}i_n}\e^{i_n{\cdots}i_{m+1}k_m{\cdots}k_1}
    \w_{k_1{\cdots}k_m}\,\j_+^{i_1}{\cdots}\j_+^{i_m}\ket{0}~,&\eInv{d}\cr
 &= \frc{f_m\,f_{n-m}}{m!}
    \e_{i_1{\cdots}i_n} ({-}1)^{{}^{n\choose2}}
      \e^{k_1{\cdots}k_mi_{m+1}{\cdots}i_n}
    \w_{k_1{\cdots}k_m}\,\j_+^{i_1}{\cdots}\j_+^{i_m}\ket{0}~,&\eInv{e}\cr
 &= \frc{f_m\,f_{n-m}}{m!}
    ({-}1)^{{}^{n\choose2}}(n{-}m)!\,
      \d^{k_1{\cdots}k_m}_{i_1{\cdots}i_m}\w_{k_1{\cdots}k_m}\,
       \j_+^{i_1}{\cdots}\j_+^{i_m}\ket{0}~,                  &\eInv{f}\cr
 &= \frc{f_m\,f_{n-m}}{m!}({-}1)^{{}^{n\choose2}}(n{-}m)!\,
      m!\,\w_{i_1{\cdots}i_m}\,
       \j_+^{i_1}{\cdots}\j_+^{i_m}\ket{0}~.                  &\eInv{g}\cr
}$$
Noting that
\eqn\eBin{ {n{-}m\choose2} ~=~
           {n\choose2} - m(n{-}m) - {m\choose2}~, }
we set
\eqn\eCoef{ f_m = {1\over m!}({-}1)^{{}^{m\choose2}}~, }
and obtain that
\eqn\eXXX{ \star\star\ket{\w;m} = ({-}1)^{m(n-m)}\ket{\w;m}~, }
\ie, that the $\star$ operator satisfies the involution axiom~\eAxi{b}.
Note also that in the complex case, $n\to2n$ and $m\to(p{+}q)$, so that
\eqn\eCpx{ ({-}1)^{m(n-m)} = ({-}1)^{(p+q)(2n-p-q)}
 =  ({-}1)^{(p+q)2n}({-}1)^{-(p+q)^2} = ({-}1)^{p+q}~, }
in perfect agreement with the standard result~\rGrHa.
\ping

Compliance with the linearity axiom, Eq.~\eAxi{c}, is obvious from the
fact that the expression~\eStAct\ for $\w^\star$ is linear in the tensor
coefficients $\w_{i_1\cdots i_m}$.

\subsec{The symmetry axiom, Eq.~\eAxi{d}}\noindent
To prove the analogue of Eq.~\eAxi{d}, we calculate:
\eqna\eSym
 $$\eqalignno{
 &\mkern-30mu\big[\a_{i_1{\cdots}i_m}\j_+^{i_1}{\cdots}\j_+^{i_m}\big]
   \Big(\star
         \big[\b_{j_1{\cdots}j_m}\j_+^{j_1}{\cdots}\j_+^{j_m}\big]\Big)\cr
 &= \big[\a_{i_1{\cdots}i_m}\j_+^{i_1}{\cdots}\j_+^{i_m}\big]\cr
 &\mkern30mu\times
   \Big[{({-}1)^{{}^{n-m\choose2}}\over(n{-}m)!}\e_{j_1{\cdots}j_n}
      \d^{j_nk_1}{\cdots}\d^{j_{n-m+1}k_m}
       \b_{k_1{\cdots}k_m}\j_+^{j_1}{\cdots}\j_+^{j_{n-m}}\Big]~,
 &\eSym{a}\cr
 &= {({-}1)^{{}^{n-m\choose2}}\over(n{-}m)!}\>
     \a_{i_1{\cdots}i_m}\,\b^{j_n{\cdots}j_{n-m+1}}\e_{j_1{\cdots}j_n}
   \j_+^{i_1}{\cdots}\j_+^{i_m} \j_+^{j_1}{\cdots}\j_+^{j_{n-m}}~,
 &\eSym{b}\cr
 &= {({-}1)^{{}^{{n\choose2}-m(n-m)-{m\choose2}}}\over(n{-}m)!}\,
     \a_{i_1{\cdots}i_m}\,
     \big[({-}1)^{m\choose2}\b^{j_{n-m+1}{\cdots}j_n}\big]\cr
 &\mkern30mu\times\big[({-}1)^{m(n-m)}
  \e_{j_{n-m+1}{\cdots}j_nj_1{\cdots}j_{n-m}}\big]
   \Big[{1\over n!}\e^{i_1{\cdots}i_mj_1{\cdots}j_{n-m}}
   \e_{\ell_1{\cdots}\ell_n}\j_+^{\ell_1}{\cdots}\j_+^{\ell_n}\Big]\,,~~
 &\eSym{c}\cr
 &= {({-}1)^{{}^{{n\choose2}}}\over(n{-}m)!}\,
     \a_{i_1{\cdots}i_m}\b^{j_{n-m+1}{\cdots}j_n}\>
  (n{-}m)!\,\d_{j_{n-m+1}{\cdots}j_n}^{~i_1~\cdots~i_m}\>
   \j_+^1{\cdots}\j_+^n~,
 &\eSym{d}\cr
 &= ({-}1)^{{}^{{n\choose2}}}\,
     \a_{i_1{\cdots}i_m}\>m!\>\b^{i_1{\cdots}i_m}\>
   \j_+^1{\cdots}\j_+^n~.
 &\eSym{e}\cr
}$$
This final form is obviously symmetric with respect to the exchange
$\a\iff\b$, verifying compliance with the symmetry axiom~\eAxi{d}.

\subsec{The non-degeneracy axiom, Eq.~\eAxi{e}}\noindent
Using the result~\eSym{e}, and substituting $\b\to\a$, we obtain that
\eqn\eSqr{ \big[\a_{i_1{\cdots}i_m}\j_+^{i_1}{\cdots}\j_+^{i_m}\big]
   \star\big[\a_{j_1{\cdots}j_m}\j_+^{j_1}{\cdots}\j_+^{j_m}\big] =
   ({-}1)^{{}^{{n\choose2}}}m!\,
     \a_{i_1{\cdots}i_m}\a^{i_1{\cdots}i_m}\>
   \j_+^1{\cdots}\j_+^n~.}
Since
\eqn\eNorm{ \a_{i_1{\cdots}i_m}\a^{i_1{\cdots}i_m} = 
 \a_{i_1{\cdots}i_m}\d^{i_1j_1}{\cdots}\d^{i_mj_m}\a_{j_1{\cdots}j_m}
 }
is (up to the combinatorial coefficient $\inv{m!}$) the standard
norm-squared of a rank-$m$ tensor, the expression~\eSqr\ vanishes only if
the tensor $\a_{i_1{\cdots}i_m}$ does. This proves non-degeneracy.

\subsec{The metric}\noindent
Using the symmetry and non-degeneracy of the product~\eSym{}, we can define
the bilinear form
\eqn\eMet{ \h(\a,\b) 
 ~\define~\int\!\!{\rm D}\f\int\rd^n\j_+
 \big[\inv{m!}\,\a_{i_1{\cdots}i_m}\j_+^{i_1}{\cdots}\j_+^{i_m}\big]
 \star\big[\inv{m!}\,\b_{j_1{\cdots}j_m}\j_+^{j_1}{\cdots}\j_+^{j_m}\big]~,
 }
which may be chosen to be the metric on the space of wave-function(al)s,
induced by the $\star$-operator.

Noting that the usual (Berezin) integration of fermions implies
\eqn\eXXX{ \int\rd^n\j_+\big[\j_+^1{\cdots}\j_+^n\big]
 ~\define~\int\rd\j_+^1{\cdots}\rd\j_+^n
  \big[({-}1)^{{}^{n\choose2}}\j_+^n{\cdots}\j_+^1\big]
 ~=~({-}1)^{{}^{n\choose2}}~, }
we obtain that
\eqn\eMetric{ \h(\a,\b) ~=~ \int\!\!{\rm D}\f~\Big(\inv{m!}\,
               \a_{i_1{\cdots}i_m}(\f)\,\b^{i_1{\cdots}i_m}(\f)\Big)~. }
Setting $\b\to\a$, we see that $\h(~,~)$ is positive definite, since the
standard norm~\eNorm\ is---as long as all the quantities involved are
real.

On the other hand, the standard field-theoretic metric on the space of
wave-function(al)s is calculated as
\eqna\eMET
 $$\eqalignno{ \V{m;\a|m;\b}
 &= \int\!\!{\rm D}\f~\big\langle0\big|
     \big[\inv{m!}\,\j_-^{i_m}{\cdots}\j_-^{i_1}\a_{i_1{\cdots}i_m}\big]
      \big[\inv{m!}\,\b_{j_1{\cdots}j_m}
       \j_+^{j_1}{\cdots}\j_+^{j_m}\big]\big|0\big\rangle~,   &\eMET{a}\cr
 &= \inv{(m!)^2}\int\!\!{\rm D}\f~\a_{i_1{\cdots}i_m}\b_{j_1{\cdots}j_m}
   \bra{0}\j_-^{i_m}{\cdots}\j_-^{i_1}\,
   \j_+^{j_1}{\cdots}\j_+^{j_m}\ket{0}~,                      &\eMET{b}\cr
 &= \inv{(m!)^2}
     \int\!\!{\rm D}\f~\a_{i_1{\cdots}i_m}\b_{j_1{\cdots}j_m}\>
       \big(m!\>\d^{i_1j_1}{\cdots}\d^{i_mj_m}\big)~,         &\eMET{c}\cr
 &= \int\!\!{\rm D}\f~
 \Big(\inv{m!}\,\a_{i_1{\cdots}i_m}(\f)\>\b^{i_1{\cdots}i_m}(\f)\Big)~.
                                                              &\eMET{d}\cr
}$$
This proves that the $\star$-induced metric~\eMet\ is the same as the
standard field-theoretic one~\eMET{}, defined using Hermitian conjugation.

Note however, that the standard formula~\eMET{} based on Hermitian
conjugation only involves (path-)integration over the bosonic degrees of
freedom, whereas the formula~\eMet\ for the $\star$-induced metric involves
an integration over both bosons and fermions. It would then appear that
this $\star$-induced formula is better suited for manifestly supersymmetric
(superfield) formulation of models which, besides simply having both
bosonic and fermionic degrees of freedom, also have supersymmetry.

\newsec{Generalizations and Summary}\noindent
The immediate consequence of the existence of the $\star$-operation is the
perfect pairing of wave-function(al)s: for every degree-$m$
wave-function(al) $\ket{\w;m}$, there exists a degree-$(n{-}m)$
wave-function(al) $\star\ket{\w;m}$. This is recognized as Poincar\'e
duality in the Hilbert space ${\cal O}_X(\wedge^*V)$. It is induced by the
Poincar\'e duality of the target manifold (coordinatized by the $\f$'s)
whenever the fermions happen to span the tangent and cotangent space to
the target manifold---as is the case in presence of supersymmetry. Even
without supersymmetry, the dynamics of the fermions may be determined by
the bosons through a $\f$-dependence of the action functional for the
fermions. In this more general case, the fermions span a bundle (or more
generally a sheaf) over the target manifold coordinatized by the bosons,
and the $\star$-operation corresponds to the Poincar\'e duality of this
bundle. The existence of the $\star$-operation is however more general:
all models with fermionic degrees of freedom have it, with or without
supersymmetry.

The recondite Reader will have observed the lack of covariance in
Eq.~\eCAR: either the $\j_+$'s and the $\j_-$'s should transform
oppositely (so that one of them should have been covariant), or the
Kronecker $\d$ should have been a metric. Indeed, following this second
option, and writing
 $$
    \big\{\, \j_-^i \,,\, \j_+^j \,\big\} ~=~ g^{ij}~, \eqno\eCAR'
 $$
leads to a simple modification of Eqs.~\eStAct\ with~\eCoef:
 $$
 \star\ket{\w;m}={(-1)^{{}^{n-m\choose2}}\over(n{-}m)!m!}
  \e_{k_1{\cdots}k_n}g^{k_nj_1}{\cdots}g^{k_{n-m+1}j_m}
   \w_{j_1{\cdots}j_m}\,\j_+^{k_1}{\cdots}\j_+^{k_{n-m}}\ket{0}~.
\eqno\eStAct'
 $$
In particular, the vanishing of all but the above shown summand in~\eStAct\
follows from the symmetry of $g^{ij}$. Were one now to vary the metric,
\ie, deform the original choice~\eCAR\ by setting $g^{ij}=\d^{ij}+h^{ij}$,
the action of the $\star$-operator on $\ket{\w,m}$ is easily seen to be of
order $m$ in the deformation $h^{ij}$.

As already noted, generalization to the complex case is straightforward.
One realizes that $n$ complex fermionic degrees of freedom of course
amounts to $2n$ real ones. The sign of the square of $\star$, given in
Eq.~\eCpx\ then agrees with the standard definition, as given for example
in Ref.~\rGrHa. The remaining calculations require little or no
correction. There appears however an interesting possibility in the
complex case; instead of the $\star$-operator~\eStar, we could now
consider the holomorphic and the antiholomorphic ``half-star''
operators\ft{On Ricci-flat manifolds, these are local expressions for the
covariantly constant holomorphic and antiholomorphic volume forms.}:
\eqna\eHSt
 $$\eqalignno{
 \Str          &\define \sum_{p=0}^n \p_p \e_{\m_1{\cdots}\m_n}\,
 \j_+^{\m_1}{\cdots}\j_+^{\m_p}\,\j_-^{\m_{p+1}}{\cdots}\j_-^{\m_n}~,
                                                              &\eHSt{a}\cr
 \overline\Str &\define \sum_{q=0}^n \bar\p_q \e_{\mb_1{\cdots}\mb_n}\,
 \j_-^{\mb_1}{\cdots}\j_-^{\mb_q}\,\j_+^{\mb_{q+1}}{\cdots}\j_+^{\mb_n}~,
                                                              &\eHSt{b}\cr
 }$$
with similarly determined (combinatorial) coefficients $\p_p,\q_q$. The
$\jb$'s in~\eHSt{b} have been reordered since the complex fermions
satisfy the hermitian canonical anticommutation relations:
\eqn\eHCAR{ \big\{\, \j_+^\m \,,\, \j_-^\nb \,\big\} ~=~ g^{\m\nb}
         ~=~ \big\{\, \j_-^\m \,,\, \j_-^\nb \,\big\}~, }
where now the $\pm$ subscripts denote ($\pm\inv2$) helicity. By defining
the $(a,c)$ Clifford-Dirac vacuum as
\eqn\eACV{ \j_-^\m\ket{0,0}_{ac} ~=~0~=~ \j_+^\mb\ket{0,0}_{ac}~. }
the annihilation operators are to the right in both of the
operators~\eHSt{}. The preceeding arguments then also prove that
they both comply with all of the axioms~\eAxi{}.

Now one can define two complex non-degenerate symmetric bilinear forms:
\eqn\eHlf{\eqalign{ h^+(\a,\b)
 &\define~\int\!\!{\rm D}\f\int\rd^n\jb_-\int\rd^n\j_+
 \big[\inv{p!}\,\a_{\m_1{\cdots}\m_p}(\f,\jb_-)
 \j_+^{\m_1}{\cdots}\j_+^{\m_p}\big]\cr
   &\mkern190mu\times
 \Str\big[\inv{p!}\,\b_{\n_1{\cdots}\n_p}(\f,\jb_-)
                      \j_+^{\n_1}{\cdots}\j_+^{\n_p}\big]~, \cr
 }}
and its $\jb_-\iff\j_+$ counterpart, $h^-(\a,\b)$. It is easy to see that
neither of these is real, as would be required of a metric. In fact, the
bilinear form defined in Eq.~\eMet\ is also complex if the quantities
involved are no longer restricted to be real. The Hermitian bilinear form,
which again turns out to be the same as the standard metric induced by
Hermitian conjugation, may be defined as:
\eqn\eMcpx{{\eqalign{g(\a,\b) 
 &\define~\int\!\!{\rm D}\f\int\rd^n\jb_-\int\rd^n\j_+
 \big[\inv{(p{+}q)!}\,\a_{\m_1{\cdots}\m_p\mb_1{\cdots}\mb_q}
 \j_+^{\m_1}{\cdots}\j_+^{\m_p}\j_-^{\mb_1}{\cdots}\j_-^{\mb_q}\big]\cr
 &\mkern210mu \times
 \star\big[\inv{(p{+}q)!}\,\bar\b_{\n_1{\cdots}\n_p\nb_1{\cdots}\nb_q}
 \j_+^{\n_1}{\cdots}\j_+^{\n_p}\j_-^{\nb_1}{\cdots}\j_-^{\nb_q}\big]~.
  }}}
This now has the usual complex linearity properties:
\eqn\eXXX{ g(\l\a,\b) = g(\a,\bar\l\b) }
The complex conjugation entered by hand in Eq.~\eMcpx\ in fact combines
the $\star$-operation with complex conjugation, the latter of which
however performed only on the bosonic coefficients. This complexified
$\star$-operation then pairs degree-$(p,q)$ wave-function(al)s with
degree-$(n{-}p,n{-}q)$ complex conjugate wave-function(al)s.
 
This complexification of the $\star$-operator could have been extended
over the fermions too, at the expense of inserting the appropriate sign
$({-}1)^{pq}$; this then would however pair degree-$(p,q)$
wave-function(al)s with degree-$(n{-}q,n{-}p)$ complex conjugate
wave-function(al)s. Of course, these two complexified
$\star$-operations simply represent the usual Poincar\'e duality with or
without complex conjugation.

The ``half-star'' operators are however more interesting, in that the
first one offers a pairing between degree-$(p,q)$ wave-function(al)s
with degree-$(n{-}p,q)$ wave-function(al)s, while its complex conjugate
pairs degree-$(p,q)$ wave-function(al)s with degree-$(p,n{-}q)$
wave-function(al)s. These turn out to represent the special holomorphic
duality on Calabi-Yau manifolds (see for example \SS\,1.2 of
Ref.~\rBeast). Note also that the above deformation formula~$\eStAct'$
simplifies when acting on the holomorphic volume form. If the deformation
is chosen to be either of pure type, $h^{ij}$ or mixed, $h^{i\bar\jmath}$,
the $\Str$-operation is now proportional to the determinant of the
deformation $h$.

The mirror map exchanges $\j_+\iff\jb_+$, whereupon we define the
$(c,c)$ Clifford-Dirac vacuum as
\eqn\eCCV{ \j_-^\m\ket{0,0}_{cc} ~=~0~=~ \j_+^\m\ket{0,0}_{cc}~. }
Interestingly, it is again possible to write down two ``twisted half-star''
operators upon rewriting $\j_\pm^\mb\to\jb_\pm^\m$:
\eqna\eMSt
 $$\eqalignno{
 \Str_-        &\define \sum_{b=0}^n f^-_b \e_{\m_1{\cdots}\m_n}\,
 \jb_-^{\m_1}{\cdots}\jb_-^{\m_b}\,\j_-^{\m_{b+1}}{\cdots}\j_-^{\m_n}~,
                                                              &\eMSt{a}\cr
 \Str_+ &\define \sum_{q=0}^n f^+_q \e_{\m_1{\cdots}\m_n}\,
 \jb_+^{\m_1}{\cdots}\jb_+^{\m_q}\,\j_+^{\m_{q+1}}{\cdots}\j_+^{\m_n}~,
                                                              &\eMSt{b}\cr
 }$$
with similarly determined (combinatorial) coefficients $f^-_b,f^+_q$. The
$\jb$'s in~\eMSt{b} have been ordered so that the annihilation operators
are to the right in both of the operators~\eMSt{}. Again, they both comply
with all of the axioms~\eAxi{}.

Now one can define two complex non-degenerate symmetric bilinear forms:
\eqna\eMlf
 $$ \eqalignno{ \m^+(\a,\b)
 &\define~\int\!\!{\rm D}\f\int\rd^n\jb_-\int\rd^n\jb_+
 \big[\inv{p!}\,\a_{\m_1{\cdots}\m_p}(\f,\jb_-)
 \jb_+^{\m_1}{\cdots}\jb_+^{\m_p}\big]\cr
   &\mkern190mu\times
 \Str\big[\inv{p!}\,\b_{\n_1{\cdots}\n_p}(\f,\jb_-)
                      \jb_+^{\n_1}{\cdots}\jb_+^{\n_p}\big]
 &\eHlf{a} \cr
 }$$
and its $\jb_-\iff\jb_+$ counterpart, $\m^-(\a,\b)$. It is easy to see that
neither of these is real, as would be required of a metric. 

The utility of the bilinear form $\h(\a,\b)$ and the mirror-pairs
of complex bilinear forms $h^\pm(\a,\b),\m^\pm(\a,\b)$ remains unclear so
far.
\ping

In conclusion, we have provided an explicit operator representation~\eStar\
of a fermionic analogue of Hodge star operation in all models with
fermionic degrees of freedom. The Hodge star operation (duality) itself
was expected since the fermions span the self-dual bundle $\wedge^*V$, the
fixed degree components of which form a Gorenstein sequence. However, no
analogue of the explicit operator representation of the Hodge star
operation is known in exterior calculus.

 This star operation induces a non-degenerate bilinear form which in the
real case is identical to the standard metric on the space of
wave-function(al)s, induced by Hermitian conjugation. In the complex case,
the star operator~\eStar\ may be factorized into its the ``half
stars''~\eHSt{}, related to the holomorphic duality on Calabi-Yau
manifolds, or the ``twisted half-stars''\eMSt{}, related to the
``half-stars'' by mirror symmetry. Furthermore, there exist several
possible complexifications of the bilinear form induced by the star,- or
the ``(twisted) half star'' operators, only one of which coincides with the
standard Hermitian metric on the space of wave-function(al)s; the utility
of the other ones remains unclear.

 %
\listrefs

\bye